\documentclass[twocolumn,prb]{revtex4}

\usepackage{graphicx}
\usepackage{dcolumn}
\usepackage{bm}

\begin{document}
\title{Kohn-Luttinger instability of the $t$-$t^\prime$ Hubbard model
in two dimensions:\\ variational approach}

\author{J. Mr\'{a}z and R. Hlubina}

\affiliation{Department of Solid State Physics, Comenius University,
Mlynsk\'{a} Dolina F2, 842 48 Bratislava, Slovakia}

\begin{abstract}
An effective Hamiltonian for the Kohn-Luttinger superconductor is
constructed and solved in the BCS approximation. The method is applied
to the $t$-$t^\prime$ Hubbard model in two dimensions with the
following results: (i) The superconducting phase diagram at half
filling is shown to provide a weak-coupling analog of the recently
proposed spin liquid state in the $J_1$-$J_2$ Heisenberg model. (ii)
In the parameter region relevant for the cuprates we have found a
nontrivial energy dependence of the gap function in the dominant
$d$-wave pairing sector. The hot spot effect in the angular dependence
of the superconducting gap is shown to be quite weak.
\end{abstract}
\pacs{PACS}
\maketitle

\section{Introduction}
In a seminal paper, Kohn and Luttinger \cite{Kohn65} (KL) presented a
perturbative argument showing that the generic ground state of a Fermi
liquid at weak coupling is superconducting, even if the bare
electron-electron interaction is not attractive in the Cooper channel
in any angular momentum sector. Recently, this result has been
rederived utilizing various versions of the renormalization group
method.\cite{Shankar94,Zanchi97,Honerkamp01}

However, no attempt has been made so far to construct a variational
wavefunction for a KL superconductor, which could be (at least in
principle) tested in a direct numerical simulation.  The purpose of
this paper is to construct such a wavefunction for the repulsive
Hubbard model in the limit of weak coupling $U\ll W$, where $U$ is the
local repulsive interaction and $W$ is the electron bandwidth. Our
strategy follows quite closely the canonical transformation approach
to the electron-phonon problem introduced by Fr\"ohlich
\cite{Frohlich52} and sketched in the Appendix.  In some sense it is
also similar to the customary analysis of the Hubbard model close to
half filling in the limit of strong coupling $U\gg W$. Namely, in the
latter case a canonical transformation can be found
\cite{Harris67,Gros87} which transforms the purely repulsive Hubbard
model to the so-called $t$-$J$ model which explicitly contains an
attractive interaction (of the order $J\sim W^2/U$) favoring the
formation of singlet superconductivity. Unfortunately, the $t$-$J$
model contains a constraint on the local number of electrons and as
such is very difficult to study. Here, on the other hand, we construct
a canonical transformation which eliminates the scattering of
quasiparticles to first order in $U/W$ and generates a weak-coupling
model with an attractive interaction of the order $U^2/W$.

The outline of the paper is as follows. The construction of the
effective model is described in Section II and our method is applied
to the two-dimensional $t$-$t^\prime$ Hubbard model in Section III.

\section{Effective Hamiltonian}
We consider the Hubbard model on a square lattice with periodic
boundary conditions and $L=l\times l$ sites, described by the
Hamiltonian $H=H_0+H_1+H_2$, where $H_0=\sum_{{\bf
k},\sigma}\varepsilon_{\bf k}n_{{\bf k},\sigma}$ is the kinetic energy
operator and the interaction term has been split into two parts:
\begin{eqnarray}
H_1&=&{U\over L}\left({N^2\over 4}-{\bf S}^2\right)
+{U\over L}\sum_{{\bf k},{\bf p}}
c^\dagger_{{\bf k} \uparrow} c^\dagger_{-{\bf k} \downarrow} 
c_{-{\bf p} \downarrow} c_{{\bf p} \uparrow},
\\
H_2&=&{U\over L}\sum_{\{{\bf k}\}}^\prime
c^\dagger_{{\bf k}_3 \uparrow} c_{{\bf k}_1 \uparrow} 
c^\dagger_{{\bf k}_4 \downarrow} c_{{\bf k}_2 \downarrow}
\delta_{{\bf k}_1+{\bf k}_2,{\bf k}_3+{\bf k}_4},
\end{eqnarray}
where $N=N_\uparrow+N_\downarrow$ and ${\bf S}$ are the total electron
number and the total spin, respectively. Terms which are not extensive
in the thermodynamic limit $L\rightarrow \infty$ have been neglected.
$H_2$ is the generic interaction term which scatters electrons from
${\bf k}_1\uparrow$ to ${\bf k}_3\uparrow$ and from ${\bf
k}_2\downarrow$ to ${\bf k}_4\downarrow$.  The prime on the summation
means that terms with ${\bf k}_1={\bf k}_3$ (forward scattering
channel), ${\bf k}_1={\bf k}_4$ (exchange channel), and ${\bf
k}_1+{\bf k}_2=0$ (Cooper channel) are excluded from $H_2$ and are
singled out into $H_1$. The operator $H_0+H_1$ can be thought of as a
reduced BCS Hamiltonian of a Landau Fermi liquid, while $H_2$ contains
the scattering processes leading to a finite lifetime of the Landau
quasiparticles.

Now we look for a canonical transformation from the bare electrons to
the dressed ones, $\tilde{H}=e^{iS}He^{-iS}$, such that the scattering
of the quasiparticles vanishes to first order in $U$.  This happens if
$H_2+i[S,H_0]=0$ in which case we can write
$\tilde{H}=H_0+H_1+i[S,H_1]+i[S,H_2]/2$, where terms of order $O(U^3)$
have been neglected.  Making use of the identity
\begin{equation}
[c^\dagger_Pc_Q,c^\dagger_Rc_S]=
\delta_{QR}c^\dagger_P c_S-\delta_{PS}c^\dagger_R c_Q
\label{eq:identity1}
\end{equation}
where the indices $P,Q,R,S$ include also the spin label, 
one verifies readily that
\begin{equation}
S={iU\over L}\sum_{\{{\bf k}\}}^\prime
{\delta_{{\bf k}_1+{\bf k}_2,{\bf k}_3+{\bf k}_4}\over
{\varepsilon_{{\bf k}_1}+\varepsilon_{{\bf k}_2}
-\varepsilon_{{\bf k}_3}-\varepsilon_{{\bf k}_4}}}
c^\dagger_{{\bf k}_3 \uparrow} c_{{\bf k}_1 \uparrow} 
c^\dagger_{{\bf k}_4 \downarrow} c_{{\bf k}_2 \downarrow}
\end{equation}
is the Hermitian generator we are looking for.  Note that the forward,
exchange, and Cooper channel processes do not contribute to $S$. In
order to calculate the commutators $[S,H_1]$ and $[S,H_2]$ we first
note that the operators $A=c^\dagger_{3\uparrow}c_{1\uparrow}$ and
$B=c^\dagger_{\gamma\uparrow} c_{\alpha\uparrow}$ commute with the
operators $X=c^\dagger_{4\downarrow} c_{2\downarrow}$ and
$Y=c^\dagger_{\delta\downarrow} c_{\beta\downarrow}$ and therefore
\begin{equation}
[AX,BY]=[A,B]XY+BA[X,Y].
\label{eq:identity2}
\end{equation}
Combining Eqs.~(\ref{eq:identity1},\ref{eq:identity2}) we thus
find the identity
\begin{widetext}
\begin{eqnarray*}
[c^\dagger_{3\uparrow} c_{1\uparrow}
c^\dagger_{4\downarrow} c_{2\downarrow},
c^\dagger_{\gamma\uparrow} c_{\alpha\uparrow}
c^\dagger_{\delta\downarrow} c_{\beta\downarrow}]=
(\delta_{1\gamma}c^\dagger_{3\uparrow} c_{\alpha\uparrow}
-\delta_{3\alpha}c^\dagger_{\gamma\uparrow} c_{1\uparrow})
c^\dagger_{4\downarrow} c_{2\downarrow}
c^\dagger_{\delta\downarrow} c_{\beta\downarrow}
+(\delta_{2\delta}c^\dagger_{4\downarrow} c_{\beta\downarrow}
-\delta_{4\beta}c^\dagger_{\delta\downarrow} c_{2\downarrow})
c^\dagger_{\gamma\uparrow} c_{\alpha\uparrow}
c^\dagger_{3\uparrow} c_{1\uparrow},
\end{eqnarray*}
making use of which one can straightforwardly calculate the effective
Hamiltonian $\tilde{H}$. The result is fairly involved, including also
three-body forces among the electrons.  Nevertheless, the expectation
value $E=\langle\tilde{\psi}|\tilde{H}|\tilde{\psi}\rangle$ in a BCS
state $|\tilde{\psi}\rangle=\Pi_{\bf k} (u_{\bf k}+v_{\bf
k}c^\dagger_{{\bf k}\uparrow} c^\dagger_{-{\bf
k}\downarrow})|0\rangle$ takes a simple and physically transparent
form,
\begin{equation}
E=\sum_{{\bf k}\sigma}\varepsilon_{\bf k}f_{{\bf k}\sigma}
+{U\over L}{N^2\over 4}
+{1\over L}\sum_{{\bf k},{\bf p}} V_{{\bf k} {\bf p}}
b^\ast_{\bf k}b_{\bf p}
+{U^2\over L^2}\sum_{\{{\bf k}\}}^\prime
{{f_{{\bf k}_1}f_{{\bf k}_2}(1-f_{{\bf k}_3})(1-f_{{\bf k}_4})}
\over
{\varepsilon_{{\bf k}_1}+\varepsilon_{{\bf k}_2}
-\varepsilon_{{\bf k}_3}-\varepsilon_{{\bf k}_4}}}
\delta_{{\bf k}_1+{\bf k}_2,{\bf k}_3+{\bf k}_4},
\label{eq:variation_en}
\end{equation}
\end{widetext}
where we have introduced a momentum-resolved BCS order parameter
$b_{\bf p}=\langle\tilde{\psi}|c_{-{\bf p}\downarrow} c_{{\bf
p}\uparrow}|\tilde{\psi}\rangle$ and we assumed
$\langle\tilde{\psi}|{\bf S}^2|\tilde{\psi}\rangle=0$. In the Cooper
channel the effective interaction reads $V_{{\bf k}{\bf p}}=U+U^2
\chi_{1}({\bf k}+{\bf p}, \varepsilon_{\bf p}-\varepsilon_{\bf k})$,
where $\chi_{1}({\bf q},\omega)$ is the real part of the particle-hole
susceptibility $\chi({\bf q},\omega)=L^{-1}\sum_{\bf K} (f_{\bf
K}-f_{{\bf K}+{\bf q}})/ (\varepsilon_{{\bf K}+{\bf
q}}-\varepsilon_{\bf K}-\omega-i0)$.  The physical content of the KL
argument is that the effective interaction between the dressed
electrons can be attractive in the Cooper channel in some angular
momentum sector. If this is the case, then the energy
$E=\langle\tilde{\psi}|e^{iS}He^{-iS}|\tilde{\psi}\rangle$ is
minimized by nonvanishing $b_{\bf p}$ and the wavefunction
$|\psi\rangle=e^{-iS}|\tilde{\psi}\rangle$ provides a variational
ansatz for the KL superconductor.

Before proceeding we would like to point out that our variational
approach bears some similarity to a very recent formulation of the
flow equation method.\cite{Grote02} The comparison of both methods is
left for future.

\section{BCS variational solution of the effective Hamiltonian}
We wish to apply the above formalism to the study of the
$t$-$t^\prime$ Hubbard model in two dimensions whose single particle
dispersion is $\varepsilon_{\bf k}=-2t(\cos k_x+\cos
k_y)+4t^\prime\cos k_x \cos k_y$, where we have set the lattice
constant $a=1$.  The superconducting phase diagram of the
weak-coupling $t$-$t^\prime$ Hubbard model has been studied previously
by considering the effective KL interaction $V^{KL}_{{\bf k}{\bf
p}}=U+U^2 \chi_{1}({\bf k}+{\bf p},0)$ with both ${\bf k}$ and ${\bf
p}$ lying at the Fermi surface.\cite{Hlubina99} Roughly speaking, such
an approach provides us with the coupling constant $g$, but not with
the prefactor $\Omega$ in the BCS formula for the transition
temperature, $T_c=\Omega\exp(-1/g)$.

In what follows we perform the standard BCS minimization of an
effective free energy whose $T=0$ limit reduces to the variational
energy Eq.~(\ref{eq:variation_en}).  We assume that the interaction
$U$ is sufficiently weak so that no particle-hole instabilities can
develop.  (In particular, this means that we can't study systems close
to the van Hove filling.)  Concentrating on lowest order effects in
$U/W$, we neglect the last term in Eq.~(\ref{eq:variation_en}) which
represents the standard second-order perturbation theory effects, and
treat it as an additive constant.  A more detailed study of
Eq.~(\ref{eq:variation_en}) taking into account also the last term is
postponed to future publications.  Moreover, since we expect that
$T_c\ll t$, we make use of the $T=0$ Fermi functions in the
calculation of the particle-hole susceptibility $\chi({\bf
q},\omega)$.

A standard BCS calculation leads to the self-consistent equation for
the gap function $\Delta_{\bf k}=L^{-1}\sum_{\bf p}V_{{\bf k}{\bf
p}}b_{\bf p}$,
\begin{equation}
\Delta_{\bf k}=-{1\over L}\sum_{\bf p}V_{{\bf k}{\bf p}}
\Delta_{\bf p}{\tanh(E_{\bf p}/2T)\over 2E_{\bf p}},
\label{eq:gap}
\end{equation}
where $E_{\bf p}=(\xi_{\bf p}^2+\Delta_{\bf p}^2)^{1/2}$, $\xi_{\bf
p}=\varepsilon_{\bf p}-\mu$, and $\mu$ is the chemical potential. 

\begin{figure}
\centerline{\includegraphics[width=10.0cm,angle=0]{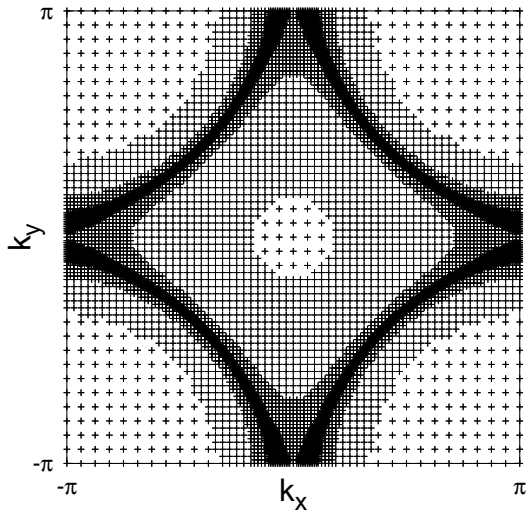}}
\caption{\label{fig:lattice} An example of the special set of ${\bf
k}$ points in the first Brillouin zone for $t^\prime/t=0.3$ and
$\rho=0.8$.  Under approaching the Fermi energy, the grid changes in
each level (there are 4 levels in total) from $32\times 32$ to
$256\times 256$.}
\end{figure}

For temperatures infinitesimally below the mean field transition
temperature, the gap equation can be linearized with respect to
$\Delta_{\bf p}$ and can be written as $D_{\bf k}=\sum_{\bf p}I_{{\bf
k}{\bf p}}(T) D_{\bf p}$, where $D_{\bf k}=\Delta_{\bf k}\phi_{\bf
k}T^{-1/2}$ is a dimensionless gap function, $\phi_{\bf
k}=[\tanh(\xi_{\bf k}/2T)/ 2\xi_{\bf k}]^{1/2}$, and $I_{{\bf k}{\bf
p}}(T)=-L^{-1}V_{{\bf k}{\bf p}} \phi_{\bf k}\phi_{\bf p}$ is a real
symmetric matrix corresponding to a dimensionless pair scattering
function. At high temperatures $I_{{\bf k}{\bf p}}(T)\propto T^{-1}$
and the gap equation does not have solutions. With decreasing $T$ the
maximal eigenvalue $\lambda$ of $I_{{\bf k}{\bf p}}(T)$ grows and the
mean field transition temperature $T_c$ can be calculated from the
condition $\lambda(T_c)=1$.  

We have studied the eigenvalues of $I_{{\bf k}{\bf p}}(T)$
numerically.  In order to minimize the cost of the calculation, the
matrix $I_{\bf k p}$ was calculated on a special lattice, see
Fig.~\ref{fig:lattice}. The lattice consists of a sequence
$i=1,2,\ldots,n$ of $n$ regular $l_i\times l_i$ grids, which become
progressively more dense as the distance to the Fermi level diminishes,
$l_{i+1}=2l_i$.  The borders between the subsequent levels are
constructed as follows: Let the maximum and minimum band energies be
$\xi_{\max}$ and $\xi_{\min}$ and let
$\xi_0=\max(\xi_{\max},|\xi_{\min}|)$.  We construct a sequence of
energies $\xi_i=q^i\xi_0$ with a quotient $q<1$.  For $i<n$, the
$i$-th level grid $l_i\times l_i$ is realized for all ${\bf k}$ points
satisfying $\xi_i<|\xi_{\bf k}|<\xi_{i-1}$, whereas the $n$-th grid
applies for $|\xi_{\bf k}|<\xi_{n-1}$.  The lattice is characterized
by three parameters: $q$, $l_1$, and the number of levels $n$. In what
follows we always take $q=4^{-1}$ and $l_1=32$, and instead of the
number of levels we describe the lattice by specifying the
finest ${\bf k}$-point grid $l_n\times l_n$.

The susceptibility $\chi({\bf q},\omega)$ was calculated on usual
lattices $l_n\times l_n$ by a straightforward modification of the
method described in Ref.~\onlinecite{Zlatic00} which makes use of the
fast Fourier transform algorithm. We have used $4l_n$ time points and
the Nyquist frequency was chosen to be $8W$.

The matrix $I_{\bf k p}$ was diagonalized by the modified Lanczos
method.\cite{Dagotto94} The initial vector was generated randomly in
one octant (quadrant) of the Brillouin zone and subsequently continued
to the whole zone so as to transform according to the one-dimensional
$s$, $d$, $d_{xy}$ and $g$ (two-dimensional $p$) irreducible
representations of the point group of the square.\cite{notation}

\begin{figure}
\centerline{\includegraphics[height=8.0cm,angle=-90]{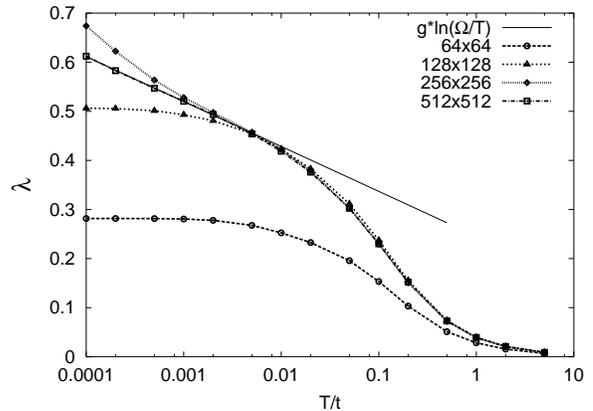}}
\caption{\label{fig:lambda_T} The temperature dependence of the
largest eigenvalue in the $d_{xy}$-wave sector for $U=W/2$,
$t^\prime/t=0.78$, and $\rho=1$. The line is a fit to
$\lambda(T)=g\log(\Omega/T)$.  The fitting parameters $g=0.04$ and
$\Omega/t=480$ imply $T_c/t=6.05\times 10^{-9}$.}
\end{figure}

As an illustrative example, the temperature dependence of the maximal
eigenvalue in the $d_{xy}$ sector for $t^\prime/t=0.78$ and electron
density $\rho=1$ is shown in Fig.~\ref{fig:lambda_T} for various
$l_n\times l_n$. Several points are worth mentioning. For a finite
lattice size $l_n$, the divergence of the particle-particle
susceptibility is cut off at temperatures $T^\star\sim v_F/l_n$, and
therefore $\lambda(T)$ saturates at low temperatures $T\ll T^\star$.
Thus, even making use of a moderate interaction
strength\cite{interaction} $U=W/2$ (this value of $U$ is used in most
numerical examples) the largest eigenvalue $\lambda$ of $I_{\bf kp}$
is typically less than unity down to the lowest
temperatures. Therefore we can't determine $T_c$ directly from
$\lambda(T_c)=1$. Instead, the $T$-dependence of the eigenvalue is
fitted in the low temperature limit by a BCS-like expression
$\lambda(T)=g\log(\Omega/T)$ and afterwards $T_c$ is calculated from
$T_c=\Omega\exp(-1/g)$. This procedure is shown explicitly in
Fig.~\ref{fig:lambda_T}. It should be pointed out that usually we can
determine the transition temperature with a reasonable degree of
confidence only if $T_c/t>10^{-6}$. In this sense the parameter values
used in Fig.~\ref{fig:lambda_T} are somewhat special.

Let us also stress that the large value of $\Omega$ estimated from
Fig.~\ref{fig:lambda_T} should not be interpreted as a large pairing
energy scale. This is evident from an equivalent but physically more
meaningful expression $\lambda(T)=\lambda(T_0)+g\log(T_0/T)$ which
makes explicit use of the pairing energy $T_0$ (its value is
arbitrary, but can be estimated from the maximal temperature where the
logarithmic scaling of $\lambda$ applies, $T_0\sim 0.005t$ in our
example).  The large value of $\Omega=T_0\exp[\lambda(T_0)/g]$ is a
consequence of the large $\lambda(T_0)$, which in turn measures the
contribution of the high energy processes to $\lambda$.

\begin{figure}
\centerline{\includegraphics[height=8.0cm,angle=-90]{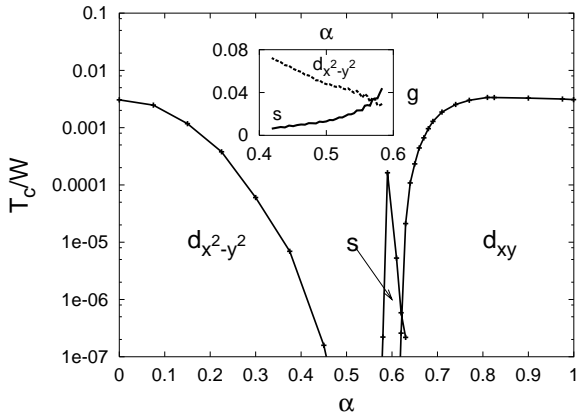}}
\caption{\label{fig:half_filling} Superconducting phase diagram for
the half filled Hubbard model at $U=W/2$. The inset shows the coupling
constant $g$ in the $d$ and $s$ sectors, calculated according to
Ref.~\onlinecite{Hlubina99} in the region where $T_c$ is not directly
accessible.}
\end{figure}

\subsection{Half-filled $t$-$t^\prime$ Hubbard model}
Now let us apply the abovementioned method to the study of the
superconducting phase diagram of the $t$-$t^\prime$ Hubbard model at
half filling ($\rho=1$).  In the strong-coupling limit $U\gg W$, this
would correspond to the Heisenberg model $H=J_1\sum_{\langle {\bf
i},{\bf j}\rangle}{\bf S_i}\cdot{\bf S_j}+ J_2\sum_{\langle\langle
{\bf i},{\bf j}\rangle\rangle} {\bf S_i}\cdot{\bf S_j},$ where
$\langle {\bf i},{\bf j}\rangle$ and $\langle\langle {\bf i},{\bf
j}\rangle\rangle$ are pairs of nearest neighbor and next-nearest
neighbor sites, respectively, $J_1=4t^2/U$, and
$J_2=4(t^\prime)^2/U$. There is general consensus that, due to
frustration, for $0.4<J_2/J_1<0.6$ the Heisenberg model does not
exhibit the N\'eel order (see, e.g. Ref.~\onlinecite{Capriotti01} and
references therein). The nature of the emergent state has been
discussed intensively, but very recently it has been
shown\cite{Capriotti01} that a projected BCS wavefunction with a mixed
$d$ and $d_{xy}$ symmetry provides a very good approximation for the
$J_1$-$J_2$ model with $J_2/J_1$ close to 1/2.

In this paper we look for a weak-coupling analogue of the phase
proposed by Capriotti {\it et al.}\cite{Capriotti01} To this end, in
Fig.~\ref{fig:half_filling} we plot $T_c$ in the $d$-wave and
$d_{xy}$-wave sectors as a function of
$\alpha=2t^\prime/(t+2t^\prime)$ for $U=W/2$ and $\rho=1$. The
dominant pairing instability is seen to change from the $d$-wave at
small $t^\prime/t$ to the $d_{xy}$ symmetry at large $t^\prime/t$,
since the susceptibility $\chi({\bf q},0)$ peaks at $(\pi,\pi)$ and
$(\pi,0)$ in these two limits.  Surprisingly, in the region
corresponding to $J_2/J_1\approx 1/2$, namely $t^\prime/t\approx
2^{-1/2}$ or $\alpha\approx 0.59$, we find that neither $d$ nor
$d_{xy}$-wave pairing instabilities are dominant, and a small island
of $s$-wave pairing is realized. For $\alpha\in[0.45,0.57]$ the
transition temperatures are too low to be accessible by the present
method. Therefore we show in Fig.~\ref{fig:half_filling} the coupling
constant $g$ in the (dominant) $d$ and $s$-wave symmetry sectors
calculated according to Ref.~\onlinecite{Hlubina99}. From the
qualitative agreement of these two different methods we conclude that
in the vicinity of $\alpha\sim 0.55$ the pairing symmetry changes from
$d$ to $s$-wave.

We expect that for $\alpha\approx 0.59$, upon increase of the
interaction strength there is presumably a true Mott-Hubbard
superconductor-insulator transition at some critical value of $U$.  We
hypothesize that before this happens, with increasing $U$ the $d$ and
$d_{xy}$ phases grow at the expense of the $s$ phase, until at some
stage the $s$ phase disappears from the phase diagram.  On the other
hand, underneath the crossing point of the $d$-wave and $d_{xy}$-wave
$T_c$'s, we expect a region with mixed $d+id_{xy}$ pairing. Such
mixed-symmetry pairing can be justified by a simple model calculation
for a two dimensional isotropic system with a separable interaction
$V(\phi,\phi^\prime)=V_1(\phi,\phi^\prime)+V_2(\phi,\phi^\prime)$ in
the Cooper channel. Here $V_1(\phi,\phi^\prime)\propto \cos 2\phi\cos
2\phi^\prime$ with coupling constant $g_1$ and cutoff $\omega_1$ leads
to $d$-wave pairing and $V_2(\phi,\phi^\prime)\propto \sin 2\phi\sin
2\phi^\prime$ with coupling constant $g_2$ and cutoff $\omega_2$
yields $d_{xy}$-wave pairing. For simplicity we further assume that by
changing a control parameter ($t^\prime/t$ in our case), only the
coupling constant $g_2$ changes. Then, if we denote that value of
$g_2$ for which the $T_c$ crossing happens as $g_{20}$, it can be
shown readily that a mixed $d+id_{xy}$ pairing state is in fact
stabilized at zero temperature in a finite window of the control
parameter for which $g_{20}^{-1}-1/2<g_2^{-1}<g_{20}^{-1}+1/2$. Note
that in the weak-coupling limit this window is quite narrow.

Before proceeding we should mention that the concept of mixed-symmetry
superconductivity has been introduced previously in the context of
strong coupling models, for the $t$-$J$ model by
Kotliar\cite{Kotliar88} and for the $t$-$t^\prime$-$J_1$-$J_2$ model
in a recent paper by Sachdev.\cite{Sachdev01} Our discussion shows
that (as pointed out already in Ref.\onlinecite{Hlubina99}),
mixed-symmetry superconductivity is in fact a generic consequence of
phase diagrams like the one shown in Fig.~\ref{fig:half_filling},
where superconducting states of different symmetry cross. 

\subsection{$d$-wave region of the $t$-$t^\prime$ Hubbard model}
\begin{figure}
\centerline{\includegraphics[width=8.0cm,angle=0]{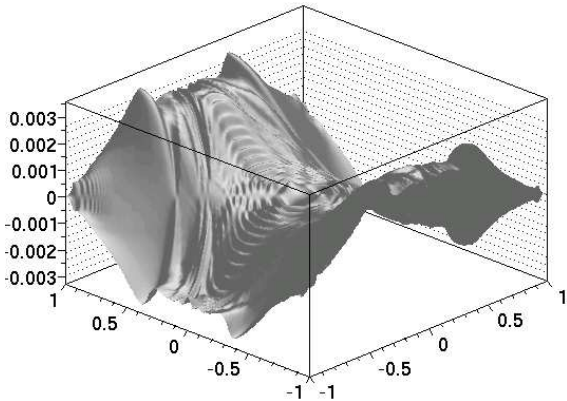}}
\caption{\label{fig:dwave_delta} The self-consistent solution
$\Delta_{\bf k}$ in the $d$-wave sector (in units of $t$) as a
function of momentum in the first Brillouin zone for $T=0$, $U=W/2$,
$t^\prime/t=0.3$, and $\rho=0.8$.  Calculated on a full lattice
$L=256\times 256$. The momenta $k_x$ and $k_y$ are measured in units
of $\pi$.}
\end{figure}

As another application of the present formalism, in the rest of this
paper we concentrate on the $d$-wave region of the superconducting
phase diagram of the $t$-$t^\prime$ Hubbard model\cite{Hlubina99}
which is relevant to the cuprates.  Close to the van Hove density,
superconductivity competes with the SDW
state\cite{Hlubina97,Hlubina99,Honerkamp01} and the question whether
one of these two states or even more exotic states\cite{Furukawa98}
wins is still open. Here we focus on a different question, namely that
of the full ${\bf k}$-dependence of the superconducting gap
$\Delta_{\bf k}$ away from the van Hove
density. Fig.~\ref{fig:dwave_delta} shows the self-consistent solution
$\Delta_{\bf k}$ of Eq.~(\ref{eq:gap}) for $t^\prime/t=0.3$,
$\rho=0.8$, and $T=0$.  The maximal gap on the Fermi surface
$\Delta_{\rm max}=0.0023t$ and the critical temperature $T_c=0.0013t$
imply $\Delta_{\rm max}/T_c\approx 1.8$.

\begin{figure}
\centerline{\includegraphics[height=8.0cm,angle=-90]{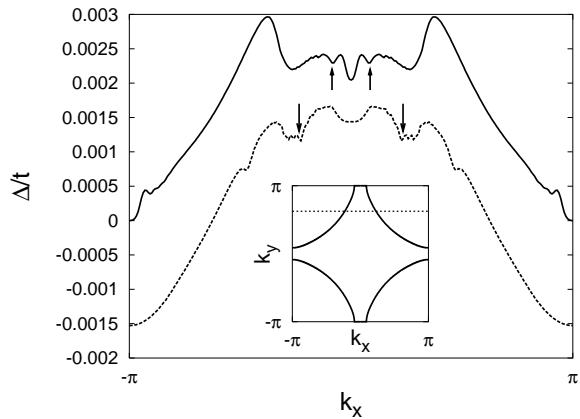}}
\caption{\label{fig:dwave_cuts} Cuts of the data in
Fig.~\ref{fig:dwave_delta} along the lines $k_y=\pi$ (top) and
$k_y=5\pi/8$ (bottom). The crossing points of the $k_y={\rm const}$
lines with the Fermi surface are indicated by arrows. The inset shows
the Fermi surface for $t^\prime/t=0.3$ and $\rho=0.8$, and the
location of the cuts.}
\end{figure}

The overall shape of $\Delta_{\bf k}$ is well described by the
simplest formula consistent with $d$-wave symmetry, $\Delta_{\bf
k}\propto \cos k_x-\cos k_y$.  Nevertheless, fine structure on top of
this overall shape is clearly visible. This is shown more
quantitatively in Fig.~\ref{fig:dwave_cuts}, where $\Delta_{\bf k}$ is
plotted along two $k_y=$const cuts of the Brillouin zone. A nontrivial
structure is seen to develop in the direction perpendicular to the
Fermi surface which is analogous to the strong-coupling effects in
conventional superconductors.

\begin{figure}
\centerline{\includegraphics[height=8.0cm,angle=-90]{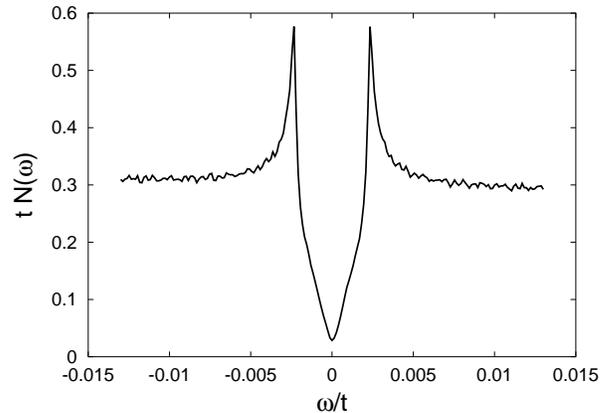}}
\caption{\label{fig:DOS_4} Tunneling density of states for $U=W/2$,
$t^\prime/t=0.3$, and $\rho=0.8$. The superconducting gap obtained on
a full lattice $L=256\times 256$ was linearly interpolated to a
lattice $L=8192\times 8192$, over which the ${\bf k}$-summation in
Eq.~\ref{eq:DOS} is taken. The delta functions appearing in the
electron spectral function Eq.~\ref{eq:spectral_f} were given a finite
width $\gamma/t=1\times 10^{-4}$.}
\end{figure}

It is well known that in the case of phonon mediated superconductivity
the energy dependence of $\Delta_{\bf k}$ leads to a structure in the
density of states. In the Appendix we have shown that our method is
capable to describe such effects. Therefore we have looked for
features in the density of states caused by the structure in
$\Delta_{\bf k}$ in our purely electronic model as well. The electron
spectral function $A({\bf k},\omega)$ and the density of states
$N(\omega)$ have been calculated from
\begin{eqnarray}
A({\bf k},\omega)&=&u_{\bf k}^2\delta(\omega-E_{\bf k})
+v_{\bf k}^2\delta(\omega+E_{\bf k}),
\label{eq:spectral_f}
\\
N(\omega)&=&L^{-1}\sum_{\bf k}A({\bf k},\omega),
\label{eq:DOS}
\end{eqnarray}
where $u_{\bf k}^2,v_{\bf k}^2=2^{-1}(1\pm\xi_{\bf k}/E_{\bf k})$.
Fig.~\ref{fig:DOS_4} shows that at weak coupling (e.g. for $U=W/2$
when $\Delta,T_c\ll t$) no structures associated with the energy
dependence of $\Delta_{\bf k}$ can be observed.  This is because
features in the energy dependence of $\Delta_{\bf k}$ can be visible
also in the density of states only if they appear in a ${\bf k}$-point
for which the single-particle energy $\xi_{\bf k}$ is not much larger
than $\Delta_{\bf k}$.  In other words, in order to be observable in
$N(\omega)$, the structure has to appear at a distance of at most
$\delta k\sim \Delta/v_F$ from the Fermi line.  At weak coupling this
criterion is not satisfied and therefore Fig.~\ref{fig:DOS_4} does not
exhibit any deviations from the textbook form of the density of states
for a $d$-wave superconductor.

\begin{figure}
\centerline{\includegraphics[height=8.0cm,angle=-90]{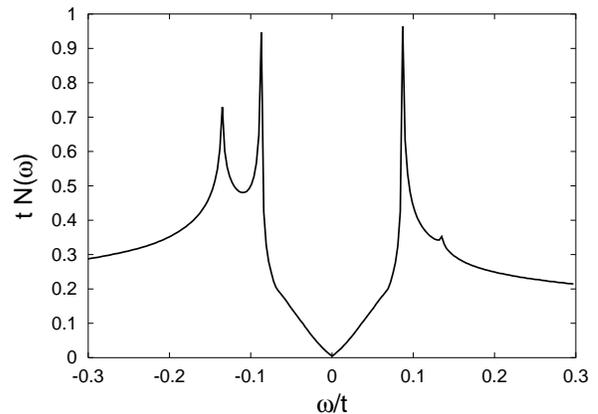}}
\caption{\label{fig:DOS_6} Tunneling density of states for $U=0.75W$,
$t^\prime/t=0.3$, and $\rho=0.8$.  Calculated by the same method as in
Fig.~\ref{fig:DOS_4} with $\gamma/t=5\times 10^{-4}$.}
\end{figure}

We have solved the self-consistent equation Eq.~(\ref{eq:gap}) also
for a larger interaction strength $U=0.75W$, when the gap increases by
more than an order of magnitude with respect to $U=W/2$.  In this case
we did find a nontrivial structure in the density of states (see
Fig.~\ref{fig:DOS_6}), which could however be attributed to the van
Hove singularity in the density of states of the noninteracting
spectrum rather than to a many body effect.

\begin{figure}
\centerline{\includegraphics[height=8.0cm,angle=-90]{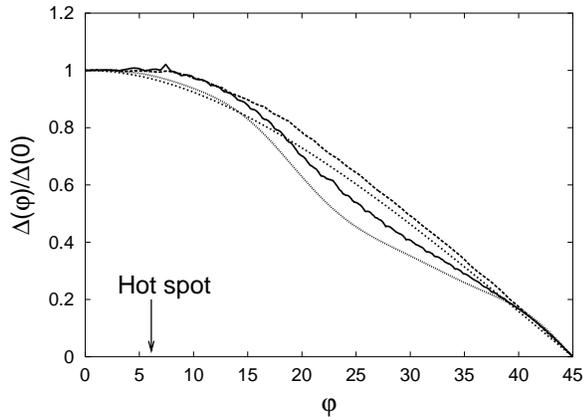}}
\caption{\label{fig:fermi_cuts} Angular variation of a normalized
superconducting gap along the Fermi line.  Calculated on a full
lattice $L=256\times 256$ for $U=W/2$, $t^\prime/t=0.3$, and
$\rho=0.8$.  Top to bottom lines at $\varphi=30^\circ$: $T/t=1$, the
simplest $d$-wave model function $\cos k_x-\cos k_y$, and $T=0$; the
lowest curve is a result obtained within the approach of
Ref.~\onlinecite{Hlubina99}.}
\end{figure}

Now let us turn to the angular dependence of $\Delta_{\bf k}$.  In
Fig.~\ref{fig:fermi_cuts} we plot $\Delta(\varphi)/\Delta(0)$ along
the Fermi line for $T=0$ and $T/t=1$. The variable $\varphi$ is the
angle between the radius vector of the Fermi surface point under study
[measured with respect to $M=(\pi,\pi)$] and the $x$ axis. The overall
shape of the high temperature data is close to the simple $\cos
k_x-\cos k_y$ form of the gap.  With decreasing temperature
$\Delta_{\bf k}$ is seen to be locally more and more enhanced close to
the crossing point of the Fermi line with the magnetic zone (hot
spot).  Nevertheless, this so-called hot spot effect is quite weak
down to $T=0$.  Fig.~\ref{fig:fermi_cuts} shows explicitly that the
hot spot effect is in qualitative agreement with an earlier
formulation \cite{Hlubina99} in which the energy dependence of the
Cooper channel interaction is neglected and $\Delta_{\bf k}$ lives
only on the Fermi line. On the other hand, a recent
paper\cite{Guinea02} reports (within the formalism of
Ref.~\onlinecite{Hlubina99}) a much stronger hot spot effect than
found here.

\section{Conclusions}
In conclusion, we have constructed a canonical transformation of the
Hubbard model which eliminates quasiparticle scattering processes to
first order in the interaction $U$. The resulting effective
Hamiltonian contains terms of order $U^2/W$ where $W$ is the
bandwidth, which are attractive in the Cooper channel.  This allows us
to construct a variational wavefunction for the Kohn-Luttinger
superconductor and to show explicitly that even a purely repulsive
model gains energy by developing superconducting correlations.

As an application of the method, the effective Hamiltonian has been
solved in the BCS approximation and the superconducting phase diagram
of the $t$-$t^\prime$ Hubbard model at half filling has been found.
The superconducting state at $t^\prime/t\sim 0.7$ is a weak-coupling
analog of the wavefunction proposed for the $J_1$-$J_2$ model by
Capriotti {\it et al.}\cite{Capriotti01} 

As another application, in the $d$-wave region of the $t$-$t^\prime$
Hubbard model \cite{Hlubina99} we have found a nontrivial energy
dependence of the gap function reminiscent of strong-coupling effects
in conventional superconductors.  There is a hot spot effect in the
angular dependence of $\Delta_{\bf k}$, which is quite consistent with
the estimates from calculations\cite{Hlubina99} in which the gap lives
only on the Fermi line.

\acknowledgments 
This work was supported by the Slovak Scientific Grant Agency under
Grant No.~VEGA-1/9177/02 and by the Slovak Science and Technology
Assistance Agency under Grant No.~APVT-51-021602.

\appendix*
\section{}
As a test of the applicability of the variational method to an
effective electronic Hamiltonian, we apply it to the case of phonon
mediated superconductivity. We consider the simplest model of
electrons with a local coupling to a phonon mode. The Hamiltonian of
the system can be written $H=H_0+H^\prime$, where
\begin{eqnarray}
H_0&=&\sum_{{\bf k},\sigma}\varepsilon_{\bf k}
c^\dagger_{{\bf k},\sigma}c_{{\bf k},\sigma}
+\sum_{{\bf q}\neq 0}\omega_{\bf q}a^\dagger_{\bf q}a_{\bf q},
\nonumber\\
H^\prime&=&{D\over \sqrt{L}}\sum_{{\bf k},\sigma,{\bf q}\neq 0}
c^\dagger_{{\bf k+q},\sigma}c_{{\bf k},\sigma}
\left(a_{\bf q}+a^\dagger_{-{\bf q}}\right).
\label{eq:model_ph}
\end{eqnarray}
Following Fr\"ohlich,\cite{Frohlich52} we seek a canonical
transformation ${\tilde H}=e^{iS}He^{-iS}$ which eliminates
electron-phonon coupling to first order in $D$.  This is accomplished
by
\begin{equation}
iS={D\over \sqrt{L}}\sum_{{\bf k},\sigma,{\bf q}\neq 0}
\left(
{c^\dagger_{{\bf k+q},\sigma}c_{{\bf k},\sigma}a_{\bf q}
\over\varepsilon_{\bf k+q}-\varepsilon_{\bf k}-\omega_{\bf q}}
+
{c^\dagger_{{\bf k+q},\sigma}c_{{\bf k},\sigma}a^\dagger_{-\bf q}
\over\varepsilon_{\bf k+q}-\varepsilon_{\bf k}+\omega_{\bf q}}
\right),
\label{eq:S_ph}
\end{equation}
since, as one verifies easily, $[iS,H_0]=-H^\prime$.  The effective
Hamiltonian can therefore be written as ${\tilde H}=H_0+{\tilde
H^\prime}$ where, to second order in $D$,
\begin{widetext}
\begin{eqnarray*}
{\tilde H^\prime}={1\over 2}[iS,H^\prime]
&=&{1\over L}\sum_{{\bf k},{\bf k^\prime},
\sigma,\sigma^\prime}\sum_{{\bf q}\neq 0}
{D^2\omega_{\bf q}\over{(\varepsilon_{\bf k+q}-\varepsilon_{\bf k})^2-
\omega_{\bf q}^2}}
c^\dagger_{{\bf k^\prime-q},\sigma^\prime}
c_{{\bf k^\prime},\sigma^\prime}
c^\dagger_{{\bf k+q},\sigma}c_{{\bf k},\sigma}
\\
&+&{D^2\over 2L}\sum_{{\bf q}\neq 0}\sum_{{\bf q^\prime}\neq 0}
\sum_{{\bf k},\sigma}
(c^\dagger_{{\bf k+q},\sigma}c_{{\bf k-q^\prime},\sigma}-
c^\dagger_{{\bf k+q+q^\prime},\sigma}c_{{\bf k},\sigma})
\left[
{a_{\bf q}(a_{\bf q^\prime}+a^\dagger_{-{\bf q^\prime}})\over
\varepsilon_{\bf k+q}-\varepsilon_{\bf k}-\omega_{\bf q}}+
{a^\dagger_{-{\bf q^\prime}}
(a_{\bf q^\prime}+a^\dagger_{-{\bf q^\prime}})\over
\varepsilon_{\bf k+q}-\varepsilon_{\bf k}-\omega_{\bf q}}
\right].
\end{eqnarray*}
The expectation value of ${\tilde H^\prime}$ in the product of a BCS
state with a phonon state diagonalizing $H_0$, $|{\tilde\psi}\rangle$,
simplifies to
$$
E=\langle\tilde\psi|{\tilde H^\prime}|{\tilde\psi}\rangle=
{1\over L}\sum_{{\bf k},{\bf k^\prime}}
V_{{\bf k},{\bf k^\prime}}b_{\bf k^\prime}^\ast b_{\bf k}
+{D^2\over L}\sum_{{\bf k},{\bf k^\prime},\sigma}
{f_{{\bf k}\sigma}(1-f_{{\bf k^\prime}\sigma})\over
\varepsilon_{\bf k}-\varepsilon_{\bf k^\prime}
-\omega_{{\bf k}-{\bf k^\prime}}}
+\sum_{{\bf q}\neq 0}\delta\omega_{\bf q}n(\omega_{\bf q}),
$$
\end{widetext}
where $n(\omega)$ is the Bose distribution function. The first term
describes the effective phonon mediated electron-electron
interaction in the Cooper channel with
$V_{{\bf k},{\bf k^\prime}}={2D^2\omega_{{\bf k}-{\bf k^\prime}}/
[(\varepsilon_{\bf k}-\varepsilon_{\bf k^\prime})^2
-\omega_{{\bf k}-{\bf k^\prime}}^2]}$, 
the second term is the standard second-order perturbation theory
correction to the ground state energy of the electron system, and the
last term with $\delta\omega_{\bf q}=-2D^2\chi_1({\bf q},\omega_{\bf
q})$ is the phonon energy renormalization.

\begin{figure}
\centerline{\includegraphics[height=8.0cm,angle=-90]{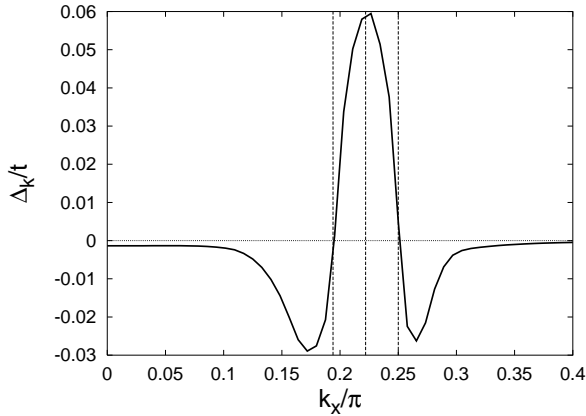}}
\caption{\label{fig:phonon} Cut of the superconducting gap for phonon
mediated superconductivity along the line $k_y=\pi$.  Calculated on a
full lattice $L=256\times 256$ for $t^\prime/t=0.45$, and $\rho=1.0$.
The phonon parameters are $\omega_0/t=0.2$ and $\Gamma/t=0.02$ and the
electron-phonon coupling $D/t=0.51$. The dashed vertical lines correspond to 
quasiparticle energies $\xi_{\bf k}=0$ and $\pm\omega_0$.}
\end{figure}

\begin{figure}
\centerline{\includegraphics[height=8.0cm,angle=-90]{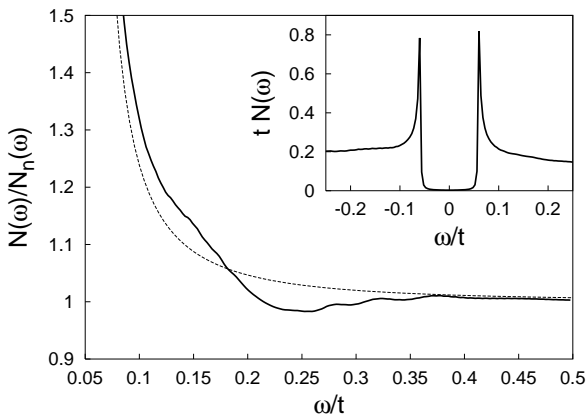}}
\caption{\label{fig:DOS_ph} Density of states in the superconducting
state $N(\omega)$ normalized by its value in the normal state
$N_n(\omega)$.  Calculated from the data in Fig.~\ref{fig:phonon}
making use of the same method as in Fig.~\ref{fig:DOS_4}. The dashed
line is the BCS approximation $N(\omega)/N_n(\omega)=
\omega/\sqrt{\omega^2-\Delta^2}$ with $\Delta=0.059t$.}
\end{figure}

In order to facilitate a numerical solution of the gap
equation Eq.~(\ref{eq:gap}) for phonon mediated pairing,
we have made use of a Cooper channel interaction  
$$
V_{{\bf k},{\bf k^\prime}}=D^2{\rm Re}\sum_{\sigma=\pm}
{\sigma\over\varepsilon_{\bf k}-\varepsilon_{\bf k^\prime}
+\sigma(\omega_{{\bf k}-{\bf k^\prime}}+i\Gamma)},
$$ regularized by a finite phonon lifetime $\Gamma$.  The BCS gap
equation has been solved numerically on the same $t$-$t^\prime$ square
lattice as for the Hubbard model.  The superconducting gap obtained
numerically for an Einstein mode with frequency $\omega_0/t=0.2$ and
$\Gamma/t=0.02$ for electronic parameters $t^\prime/t=0.45$ and
$\rho=1.0$ and electron-phonon coupling $D/t=0.51$ is shown in
Fig.~\ref{fig:phonon}. Note that the gap changes sign in the vicinity
of the phonon energy $\omega_0$. This nontrivial energy dependence of
$\Delta_{\bf k}$ translates into a feature of the tunneling density of
states in the same energy region, see Fig.~\ref{fig:DOS_ph}. It should
be pointed out that although both of the above results are in
qualitative agreement with the Eliashberg theory,\cite{Scalapino69}
there is one important difference of the present approach with respect
to Ref.~\onlinecite{Scalapino69}. Namely, within the Eliashberg theory
the features of $\Delta(\omega)$ and $N(\omega)$ occur at energy
$\omega_0+\Delta(0)$ and not $\omega_0$ as in our case. This
difference is presumably due to the fact that within the canonical
transformation approach, it is the bare electron energies which enter
the energy denominators in Eq.~(\ref{eq:S_ph}) and not the
renormalized energies as in the Eliashberg theory. However, in a truly
weak coupling theory this difference is marginal.


\begin{thebibliography}{99}
\bibitem{notation}The notation is the same as in
  Ref.~\onlinecite{Hlubina99}.
\end{thebibliography}
\end{document}